\documentclass[12pt]{article} \usepackage{graphicx}
\usepackage{color}
\usepackage{colortbl}
\usepackage{rotating}
\usepackage{amsmath} 
\usepackage{ifthen} 
\usepackage{multirow}
\usepackage{subfigure}
\usepackage{rotating}
\usepackage{placeins}
\usepackage{morefloats}
\usepackage{amssymb}
\usepackage{amsfonts}
\usepackage{upgreek} 
\usepackage{xspace}

\usepackage{hyperref}    
\usepackage[all]{hypcap} 

\def\lhcb {LHCb\xspace}
\def\babar  {BaBar\xspace}
\def\belle  {Belle\xspace}
\def\cdf    {CDF\xspace}
\def\cleo   {CLEO\xspace}
\def\PK      {\ensuremath{\mathrm{K}}\xspace}                 
\def\PD      {\ensuremath{\mathrm{D}}\xspace}                 
\def\Ppi         {\ensuremath{\uppi}\xspace}

\def\kaon  {\ensuremath{\PK}\xspace}

\def\Kmp    {\ensuremath{\kaon^-\kaon^+}\xspace}
\def\pion  {\ensuremath{\Ppi}\xspace}

\def\pimp   {\ensuremath{\pion^-\pion^+}\xspace}
\def\Dstar   {\ensuremath{$D$^*}\xspace}
\def\Dbar    {\kern 0.2em\overline{\kern -0.2em \PD}{}\xspace}
\def\D       {\ensuremath{D}\xspace}

\def\Dz      {\ensuremath{\mbox{D}^0}\xspace}
\def\Dzb     {\ensuremath{\Dbar^0}\xspace}
\def\Dznospace      {\ensuremath{\mbox{D}^0}}

\def\DzDzb   {\ensuremath{\Dz {\kern -0.16em \Dzb}}\xspace}
\def\CP                {\ensuremath{C\!P}\xspace}

\def\ycp        {\ensuremath{y_{CP}}}
\def\agamma     {\ensuremath{A_{\Gamma}}}
\def\kpi        {\ensuremath{\PK\Ppi}\xspace}
\def\kk         {\ensuremath{\PK\PK}\xspace}
\newcommand{\decay}[2]{\ensuremath{#1\!\to #2}\xspace}         
\def\kpi        {\ensuremath{\PK\Ppi}\xspace}
\def\pipi        {\ensuremath{\Ppi\Ppi}\xspace}
\def\kk         {\ensuremath{\PK\PK}\xspace}

\def\dpipic       {\decay{\Dz}{\pimp}}
\def\dkkc        {\decay{\Dz}{\Kmp}}

\def\pbnr{} \def\speaker{Silvia Borghi}
\def\title{Indirect \CP violation results and HFAG averages}
\def\affiliation{School of Physics and Astronomy\\ The University of
  Manchester, Manchester, UK}

\newcommand\pubnumber{\pbnr}
\newcommand\pubdate{\today}
\def\Title#1{\begin{center} {\Large #1 } \end{center}}
\def\Author#1{\begin{center}{ \sc #1} \end{center}}

\newcommand{\OnBehalf}[1]{\sbox0{#1}\ifdim\wd0=0pt
        {}
	\else
	{\\on behalf of #1}
	\fi}
\newcommand{\SupportedBy}[1]{\sbox0{#1}\ifdim\wd0=0pt
        {}
	\else
	{\footnote{#1}}
	\fi}
\def\Address#1{\begin{center}{ \it #1} \end{center}}

\newcommand\pubblock{\includegraphics[width=5cm]{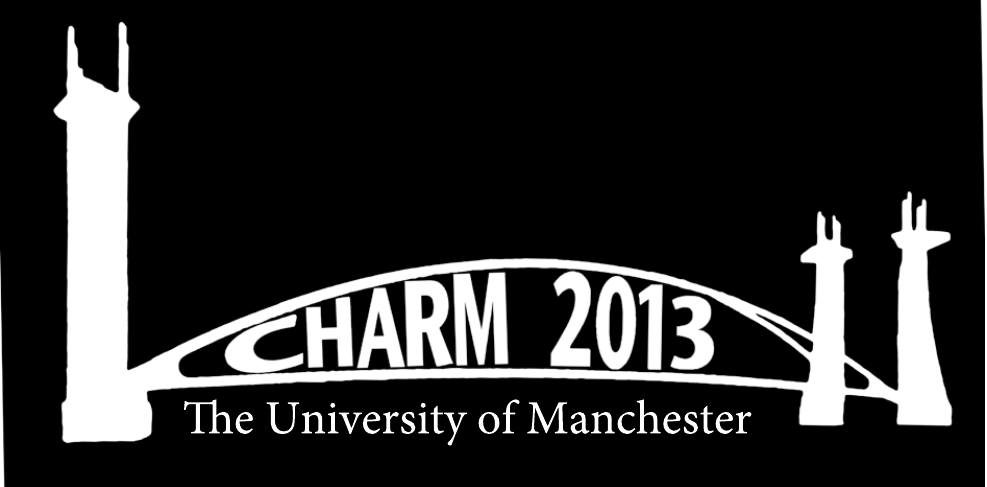}\hfill{\begin{tabular}{l} \pubnumber\\
         \pubdate  \end{tabular}}}
\newenvironment{Abstract}{\begin{quotation}  }{\end{quotation}}
\newenvironment{Presented}{\begin{quotation} \begin{center} 
             PRESENTED AT\end{center}\bigskip 
      \begin{center}\begin{large}}{\end{large}\end{center} \end{quotation}}

\def\venue{The 6$^{th}$ International Workshop on Charm Physics\\
(CHARM 2013)\\
Manchester, UK,  31 August -- 4 September, 2013}

\def\beq{\begin{equation}}
\def\eeq#1{\label{#1}\end{equation}}
\def\eeqn{\end{equation}}

\def\beqa{\begin{eqnarray}}
\def\eeqa#1{\label{#1}\end{eqnarray}}
\def\eeqan{\end{eqnarray}}

\let\bar=\overbar

\def\D{{\cal D}}

\def\Dslash{\not{\hbox{\kern-4pt $D$}}}
\def\dslash{\not{\hbox{\kern-2pt $\del$}}}

\def\msb{{\bar{\ssstyle M \kern -1pt S}}}

\begin{document}
\begin{titlepage}
\pubblock

\vfill \Title{\title} \vfill
\Author{\speaker} \Address{\affiliation} \vfill
\begin{Abstract}
The current status of the search for indirect \CP
violation in the neutral D meson system at the B-factories and at \lhcb is reported.
The indirect \CP asymmetry search is performed by the measurement of the 
proper-time asymmetry (\agamma)
in decays of \Dz \Dzb mesons to  \CP eigenstates, \Kmp and \pimp, 
and by \ycp, the ratio between the effective lifetime 
measured in decay to a \CP eigenstate and that to the mixed eigenstate \kpi.
All results are consistent with the no \CP violation hypothesis.
The latest world averages for mixing and \CP asymmetry in the charm sector
 evaluated by the Heavy Flavour Averaging Group are presented. 
The no mixing hypothesis  is excluded at more than 12 standard deviations.
The search for direct and indirect \CP violation in the  charm sector is consistent with
no \CP violation at 2.0\% confident level.

\end{Abstract}
\vfill
\begin{Presented}
\venue
\end{Presented}
\vfill
\end{titlepage}
\def\thefootnote{\fnsymbol{footnote}} \setcounter{footnote}{0}
%

\section{Introduction}
The charm sector is a promising field to probe for the effect of physics 
beyond the Standard Model (SM).
The study of neutral \D~mesons offers a unique opportunity to access up-type 
quarks in flavour-changing neutral current (FCNC) processes. 
In the SM, indirect \CP violation in the charm sector is 
expected to be highly suppressed, less than $\mathcal{O}(10^{-3})$, and 
at first order independent of the final state.  
Direct \CP violation can be larger in SM and depends 
on the final state \cite{theorysm}.  Both asymmetries can be enhanced by New
Physics in principle up to  $\mathcal{O}(1\%)$ \cite{theoryNP,theory1,theory2}.
The charm-mixing process has recently been observed for the first time 
unambiguously in single measurements \cite{lhcbmixing1,lhcbmixing2}.
Charge Parity (\CP) violation, on the other hand, has yet not been observed.
Evidences of direct \CP violation by measuring the difference of 
\CP asymmetries in singly-Cabibbo suppressed two-body decays have been reported by the \lhcb and \cdf 
experiments \cite{acpfirst,acpcdf}, but it
has not been confirmed by the most recent measurements 
\cite{acpsecond,acpsemilept}.

Flavour mixing occurs when the mass eigenstates ($|\mbox{D} _{1,2} \rangle$) differ from 
the flavour eigenstates and they can be written as linear combinations of the 
flavour eigenstates $|\mbox{D}_{1,2} \rangle=p|\Dz \rangle\pm{}q|\Dzb \rangle$, with complex 
coefficients $p$ and $q$ which satisfy $|p|^2+|q|^2=1$.
The mixing parameters are defined as $x
\equiv (m_1-m_2)/ \Gamma $ and 
$y\equiv (\Gamma_1-\Gamma_2)/(2\Gamma)$, where $m_1$, $m_2$, $\Gamma_1$ and 
$\Gamma_2$ are the masses and the decay widths for $\mbox{D}_{1}$ and $\mbox{D}_{2}$, respectively, 
and $\Gamma=(\Gamma_1+\Gamma_2)/2$.
The phase convention is chosen such that $CP |\Dz \rangle=-|\Dzb\rangle$.
In the absence 
of \CP violation $q/p=1$, $\mbox{D}_{1}$ is \CP-even and $\mbox{D}_{2}$ is \CP-odd.

Three types of \CP violating effects can be distinguished. Firstly \CP violation in decays 
occurs when the decay amplitude differs for particle and anti-particle: 
$|\bar{A}_f/A_f| = 1 + A_d$  where $A_d$ is the contribution from direct \CP 
asymmetry.
Secondly, \CP violation in mixing occurs when $q$ and $p$ differ ($| q/p| = 1 \pm A_m$ where $A_m$ is the mixing \CP contribution).
Finally, \CP violation contribution appears due to the interference between mixing 
and decay amplitudes when $\phi \equiv arg [q\bar{A}_f/pA_f]$.
The \CP direct and indirect asymmetries can be defined as 
\begin{equation}\label{eqdir}
a_{\CP}^{dir} \equiv -1/2 \: A_d
\end{equation}
\begin{equation}\label{eqindir}
a_{\CP}^{ind} \equiv - A_m/2 \: y \cos \phi + x \sin \phi .
\end{equation}

From the study of the lifetime of \CP eigenstates
and mixed eigenstates one can extract information about the mixing and
\CP violation as they  would modify the decay time distribution of \CP eigenstates.  
One of the interesting parameters is \ycp~which is
defined as the ratio of the lifetime of  the mixed eigenstate 
and a \CP eigenstate where the final state ($f$) is $K^-K^+$ or $\pi^-\pi^+$,
$$
y_{CP}= \frac{\tau  \left( \Dznospace  \rightarrow K^-\pi^+\right )}
            {\tau\left(\Dznospace \rightarrow f\right)}
-1 \cong \left(1-\frac{1}{8}A_m^2\right) y \cos \phi-\frac{1}{2}A_m x \sin \phi.
$$

In case of no \CP violation \ycp~is equal to the
$y$ mixing parameter. On the other hand, if \ycp~differs from the 
$y$ this is a sign of \CP violation.

The other interesting parameter that one can extract from the lifetime
measurements of \CP eigenstates is \agamma~defined as
$$
A_{\Gamma}=\frac{\tau  (\Dzb \rightarrow f)-\tau (\Dz \rightarrow f)}
                 {\tau(\Dzb \rightarrow f)+\tau (\Dz \rightarrow f)}
\approx \frac{1}{2} \left(A_m+A_d\right) y  \cos \phi - x \sin \phi A_m 
$$

A measurement of $A_{\Gamma}$ differing significantly from zero is a 
manifestation of \CP violation as it requires a non-zero value
for $A_m$, $A_d$ or $\phi$.
\babar determines the parameter $\Delta$Y instead of \agamma~which is written as
$\Delta$Y=$\left( 1+\ycp \right) \agamma$. 
The SM predicts \agamma~to be smaller than $10^{-4}$ while physics
beyond the SM estimate \agamma~up to $10^{-2}$. 

The difference in the two final states ($K^-K^+$, $\pi^-\pi^+$) can be evaluated as
$$
\Delta A_{\Gamma}\left(f(KK)-f(\pi\pi)\right) \approx \Delta A_d y \cos \phi + \left( A_m+A_d \right) y \Delta \cos \phi - x \Delta \sin \phi
$$
Assuming both $x$ and $y$ of $\mathcal{O}(10^{-2})$, $\cos\phi=1$ and direct \CP violation at the level of $A_d/2 \approx 1\%$
would lead to a difference of the per mille.
The B-factories neglect terms of $\mathcal{O}(10^{-4})$, hence they do not consider any difference between the final states.
While \lhcb quotes the results for the two final states, $\mbox{K}^-\mbox{K}^+$ and $\pi^-\pi^+$, separately.

\section{Experimental measurements}
The measurement of \agamma~requires the flavour tagging of the \Dz-\Dzb.
This is obtained using a \Dstar sample where the \Dstar is  promptly decaying
at the primary vertex into \Dz and a charged pion. 
The charge of the pion (usually called slow pion due to its low momentum compared 
to the pions of \Dz daughters) determines univocally the \Dz flavour.

Several types of background are found in \Dstar samples:
combinatorial background, specific background coming from mis- or partially 
reconstructed events, and association of a true \Dz with a random slow pion.
The combinatorial background can be determined using the side band of
the \Dz invariant mass. While the decays that could contribute to 
the specific background are studied deeply using simulated samples
to determine the shape in the mass distribution, the yields are determined
in the fit procedure.
The latest type of background  can be studied using 
$\Delta m$ defined as the difference of invariant mass of 
\Dstar and \Dz.

\subsection{Measurements at the B-factories}

The \babar and \belle experiments use a similar procedure to determine the \CP parameters.
After a selection based on track and vertex quality, on particle identification (PID) and
on variables that remove \D~candidates originating B decays,
an unbinned maximum likelihood fit
of \Dz invariant mass determines the background and optimizes the selection on the \Dz invariant mass and $\Delta m$.
Purities better than 98\%
for \Kmp and 93\% for \pimp are achieved. 
Both experiments  use the full data sample collected.
The total data sample contains a few hundred thousand
candidates for \kk, about a hundred thousand for \pipi and few million candidates for the Cabibbo
favoured mode \kpi.

At \babar, the effective lifetimes and the parameters \ycp~and $\Delta$Y are evaluated by
simultaneous fit of seven samples: five tagged samples and two untagged samples for \kk and \kpi.
The results are \cite{babar}:
\begin{eqnarray}
\ycp & = & (0.72\pm0.18_{stat}\pm0.12_{syst})\% \nonumber\\
\Delta Y  & = & (0.09\pm0.26_{stat}\pm0.06_{syst})\% \nonumber\\
\end{eqnarray}
The result of \ycp~shows an
evidence for mixing at 3.3~$\sigma$ and the $\Delta$Y parameter is compatible with 0.  Thus no \CP asymmetry is
observed.

At \belle experiment a simultaneous binned maximum likelihood fit to
all 5 tagged samples is performed to extract the effective
lifetimes, \ycp~and \agamma.
The data are divided into two independent data sets due to the
different vertex detector used during data taking.  
In addition it has been
found that the \Dz lifetime has a dependence on the \Dz polar angle in
the central mass frame due to dependence of the lifetime resolution
function.  Thus it is needed to perform the fit in bins of polar angle ($\theta$).
The weighted average of these measurements results to be \cite{belle}:

\begin{eqnarray}
\ycp & = & (1.11\pm0.22_{stat}\pm0.11_{syst})\% \nonumber\\
\agamma  & = & (0.03\pm0.20_{stat}\pm0.08_{syst})\% \nonumber\\
\end{eqnarray}

The \ycp~measurement shows a  mixing at 4.5$\sigma$. No sign of \CP
violation is observed as the value of \agamma~is compatible with zero.

\subsection{The measurements at \lhcb}
The measurements at \lhcb are performed using 1~fb$^{-1}$ of data from a sample
of $pp$ collisions at a centre-of-mass energy of 7 TeV collected in 2011. The main selection is applied at the trigger level on the momentum, PID and impact parameter (IP) of the \Dz daughters. 
Only few additional offline requirements, e.g. on the track quality and on the $\Delta$m window, are applied.

The data sample contains almost five millions of \Kmp candidates and 1 million of \pimp. 
A purity of 93.6\% (91.2 \%) for \kk (\pipi) is obtained in a region of two standard deviations of the signal peak and $\Delta$m.
The data sample is splitted into four independent sets dependending on the magnet polarity and on two run periods due to different detector alignment and calibration conditions.
The fit to determine the effective lifetime is performed independently for each data set, each flavour tag and each decay mode.  

\begin{figure}[!tb]
\centering
\includegraphics[height=2.2in]{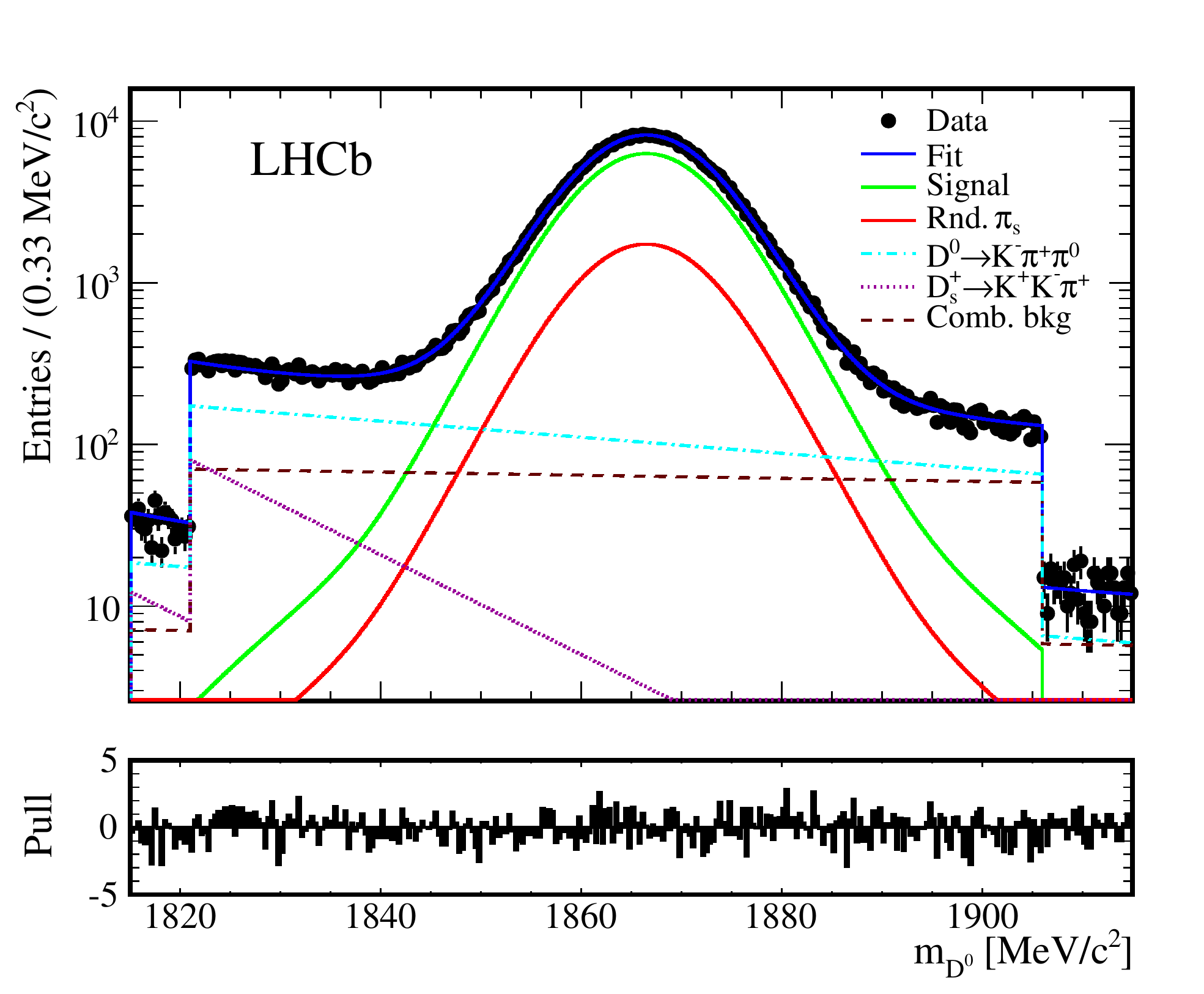}
\includegraphics[height=2.2in]{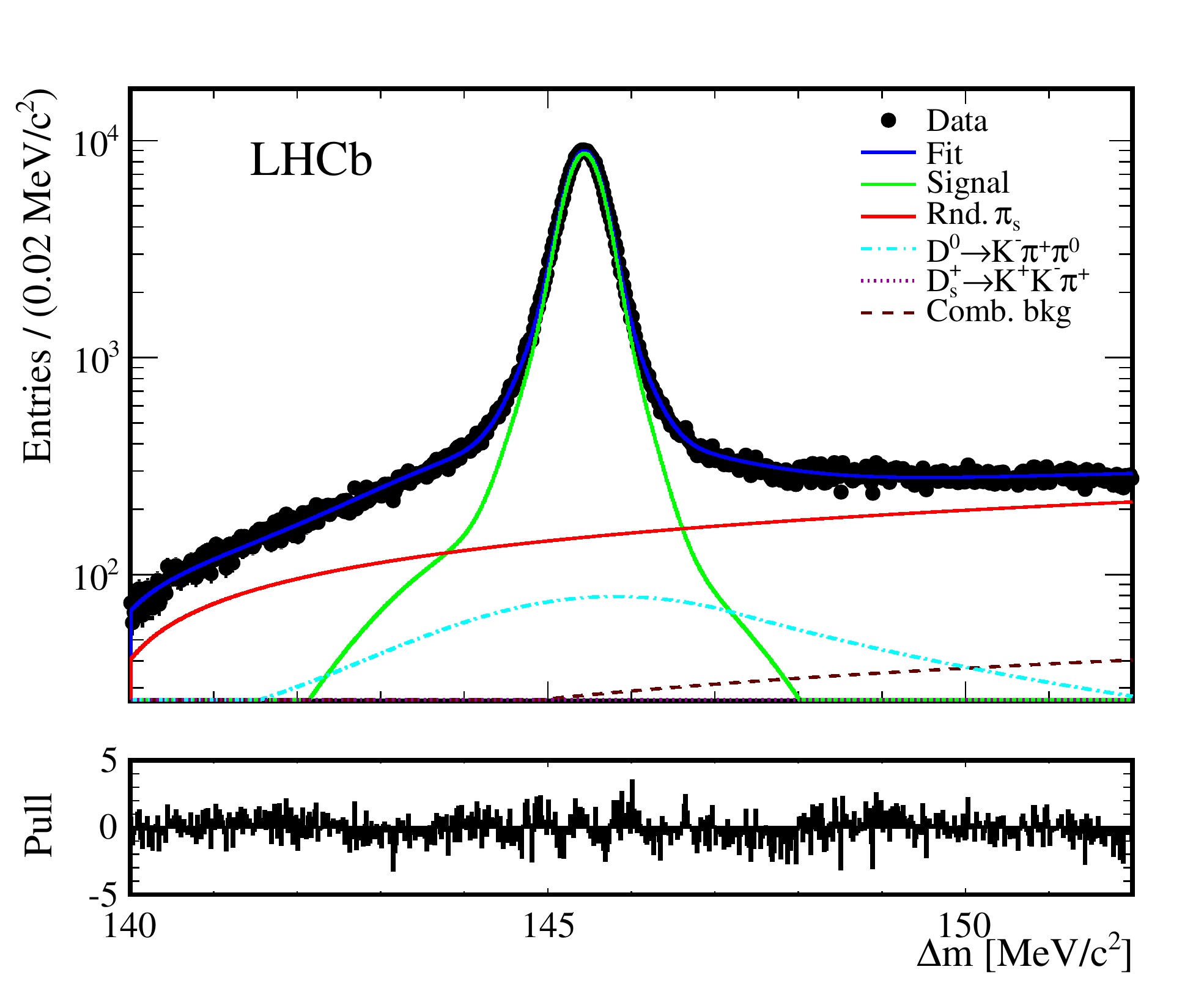}
\caption{On the left, the \Dz invariant mass and on the right the $\Delta$m mass difference between the reconstructed \Dstar and \Dz candidates. The data are shown as points.}
\label{fig:mass}
\end{figure}

The signal yield and the background contribution are extracted from  simultaneous unbinned likelihoood 
fits of the \Dz invariant mass and $\Delta$m that allow to distinguish the different background sources. 
One example of the mass fit results is shown in Fig. \ref{fig:mass}.
The only background that is not distinguishable by the mass fit is the one due to the secondary charm.
The secondaries have larger impact
parameter with respect to the primary vertex than the prompt candidates as a secondary $D$ does not usually
 point back to the primary vertex.  
Thus  this background can be reduced by a selection  based on the topology
but it can not be completely suppressed. The variable
 $\mbox{ln} \left( \chi^2_{IP} \right)$ is used.
It is defined as the difference in $\chi^2$
of a given primary interaction vertex reconstructed with and without the considered particle.

\begin{figure}[!tb]
\centering
\includegraphics[height=2.2in]{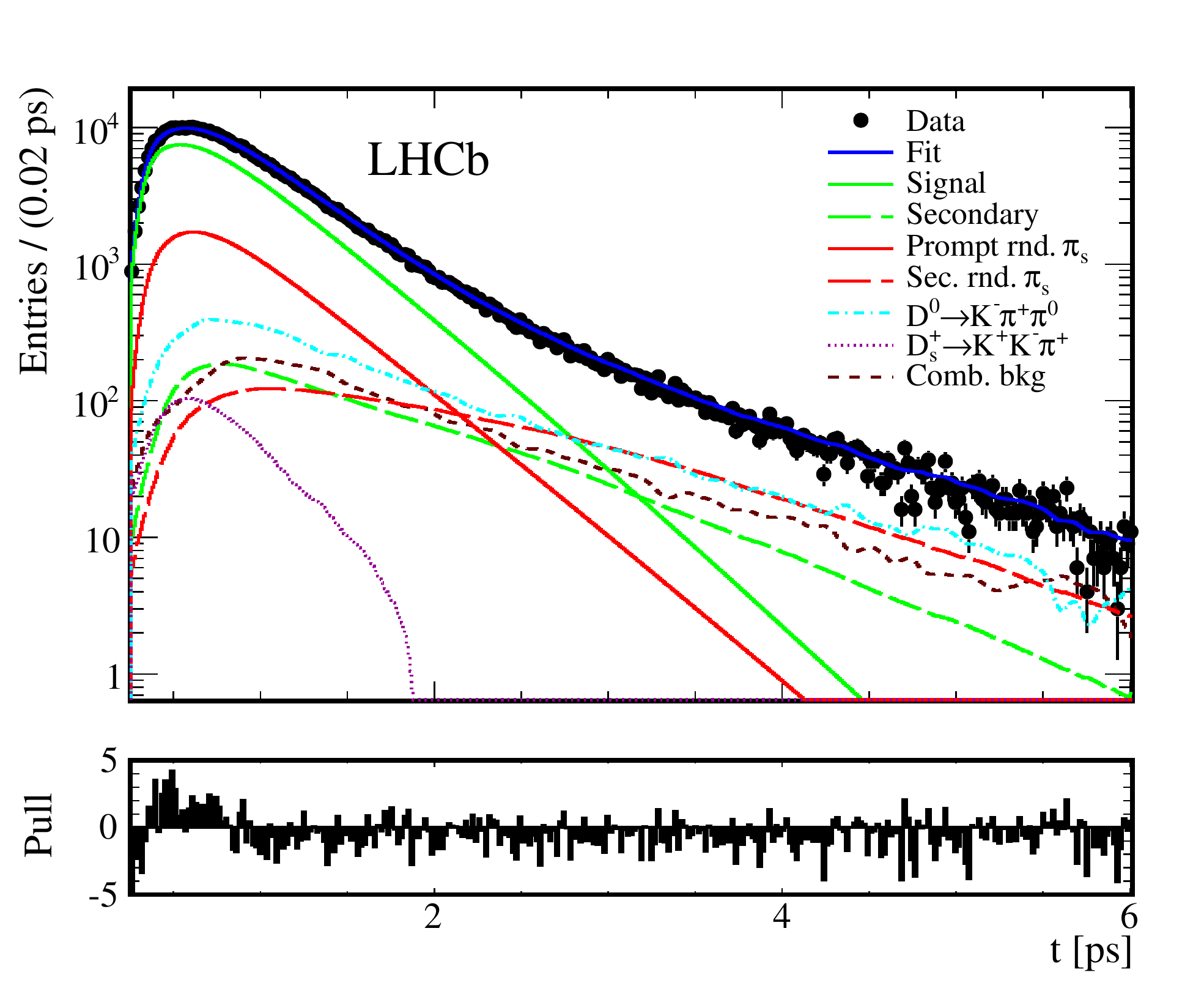}
\includegraphics[height=2.2in]{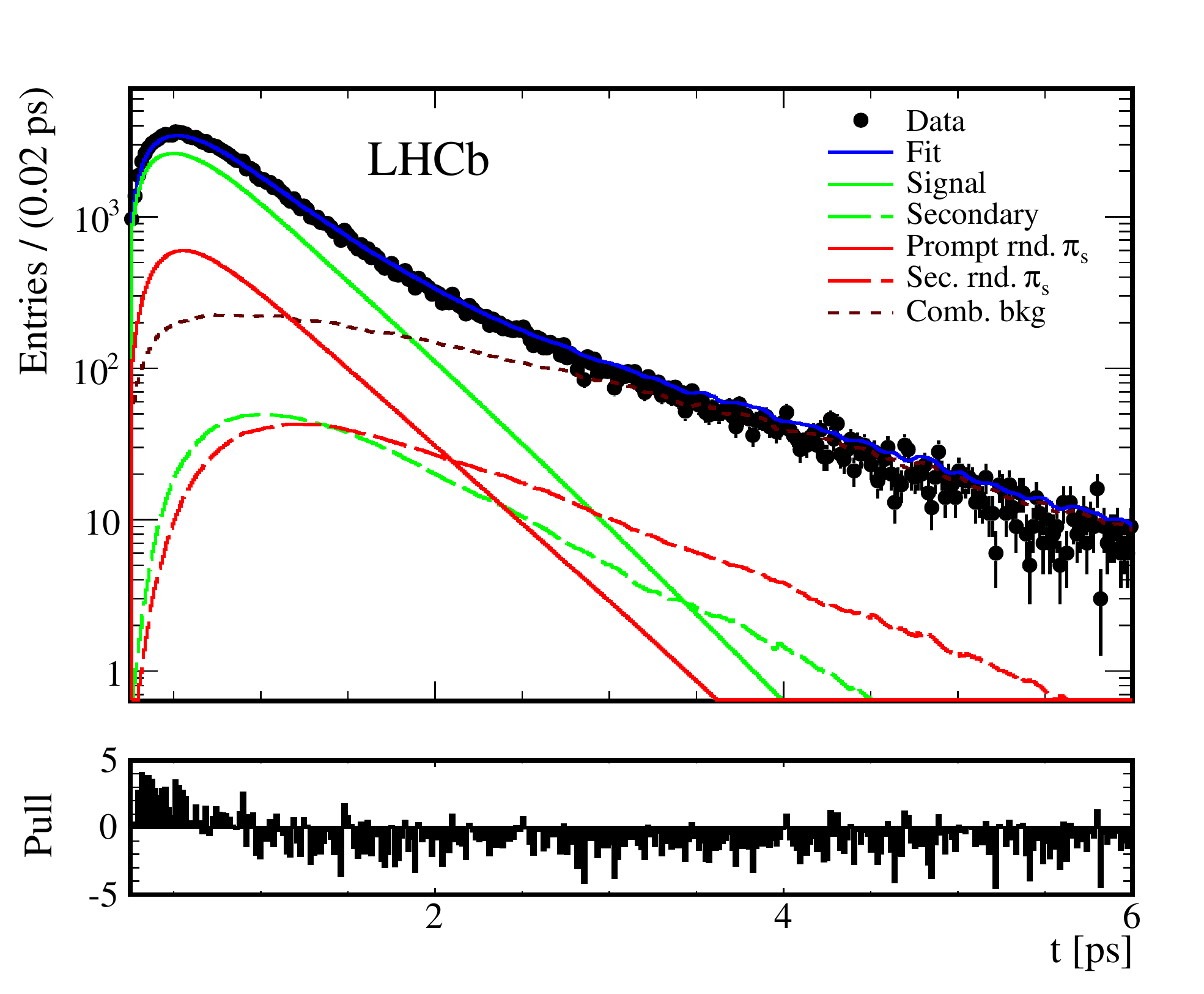}
\caption{Lifetime fit projection of  \dkkc (on the left) and \dpipic (on the right) candidates. The dare are shown as points.}
\label{fig:timeft}
\end{figure}

The measurement of $A_{\Gamma}$ is performed by the 
measurements of the effective lifetime obtained by a simultaneous fit of proper time and  $\mbox{ln}
\left( \chi^2_{IP} \right)$. 
For the evaluation of the effective lifetime an acceptance correction is applied because of the bias of the measured proper-time distribution due to the trigger event selection. The acceptance is determined using a data driven method, the so-called swimming algorithm \cite{swimming,swimming2,swimming3,agammafirst}.
Figure \ref{fig:timeft} shows an example of the lifetime projection 
for $\Dz \rightarrow K^-K^+$ and $\Dz \rightarrow \pi^-\pi^+$ decays. 

The method is validated on a control measurement using decays to the
Cabibbo favoured decay $ \Dz \rightarrow K \pi$. 
The lifetime asymmetry is determined to be consistent with zero in accordance with the 
expectation.

\begin{figure}[!tb]
\centering
\includegraphics[height=2.2in]{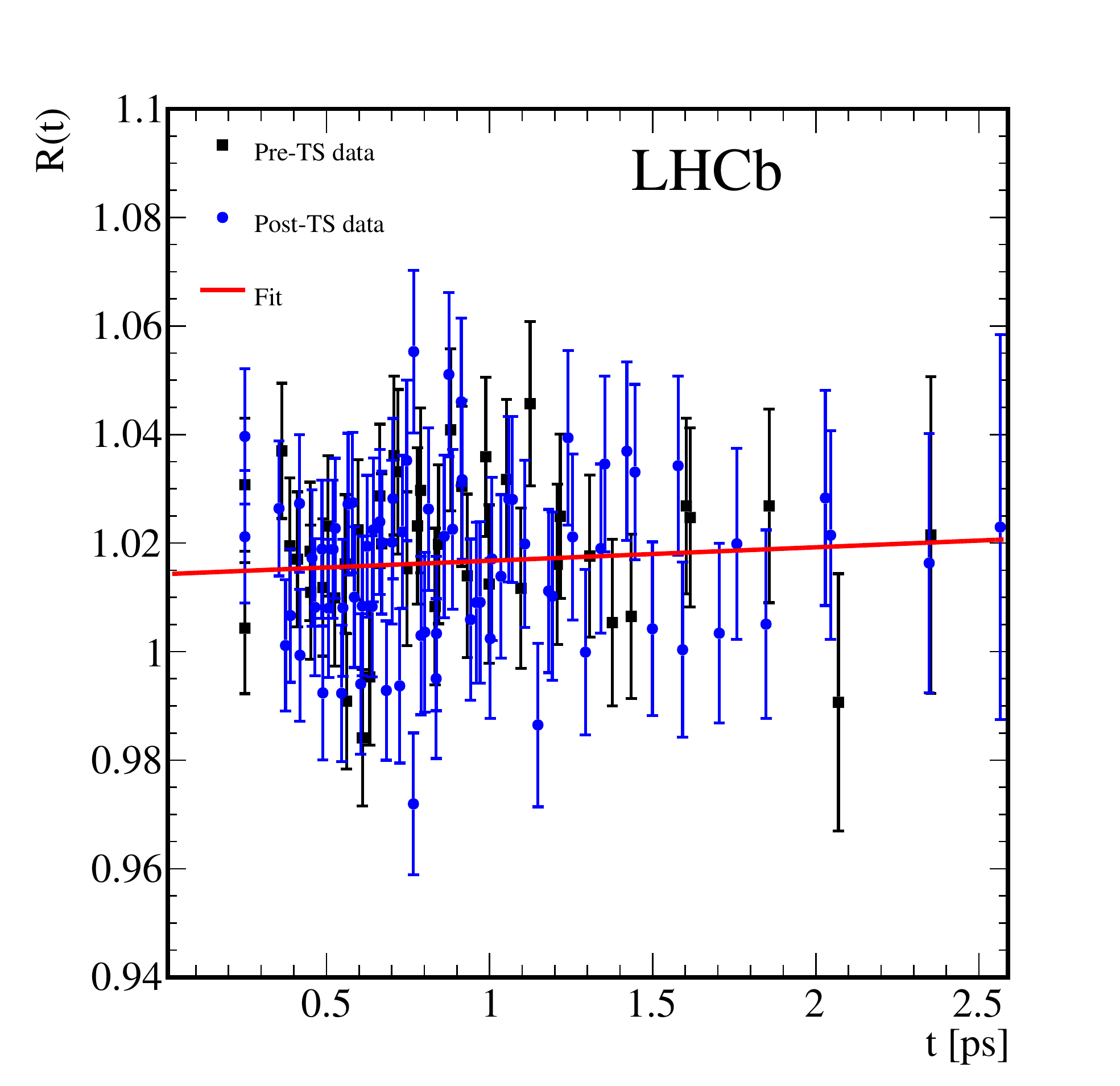}
\includegraphics[height=2.2in]{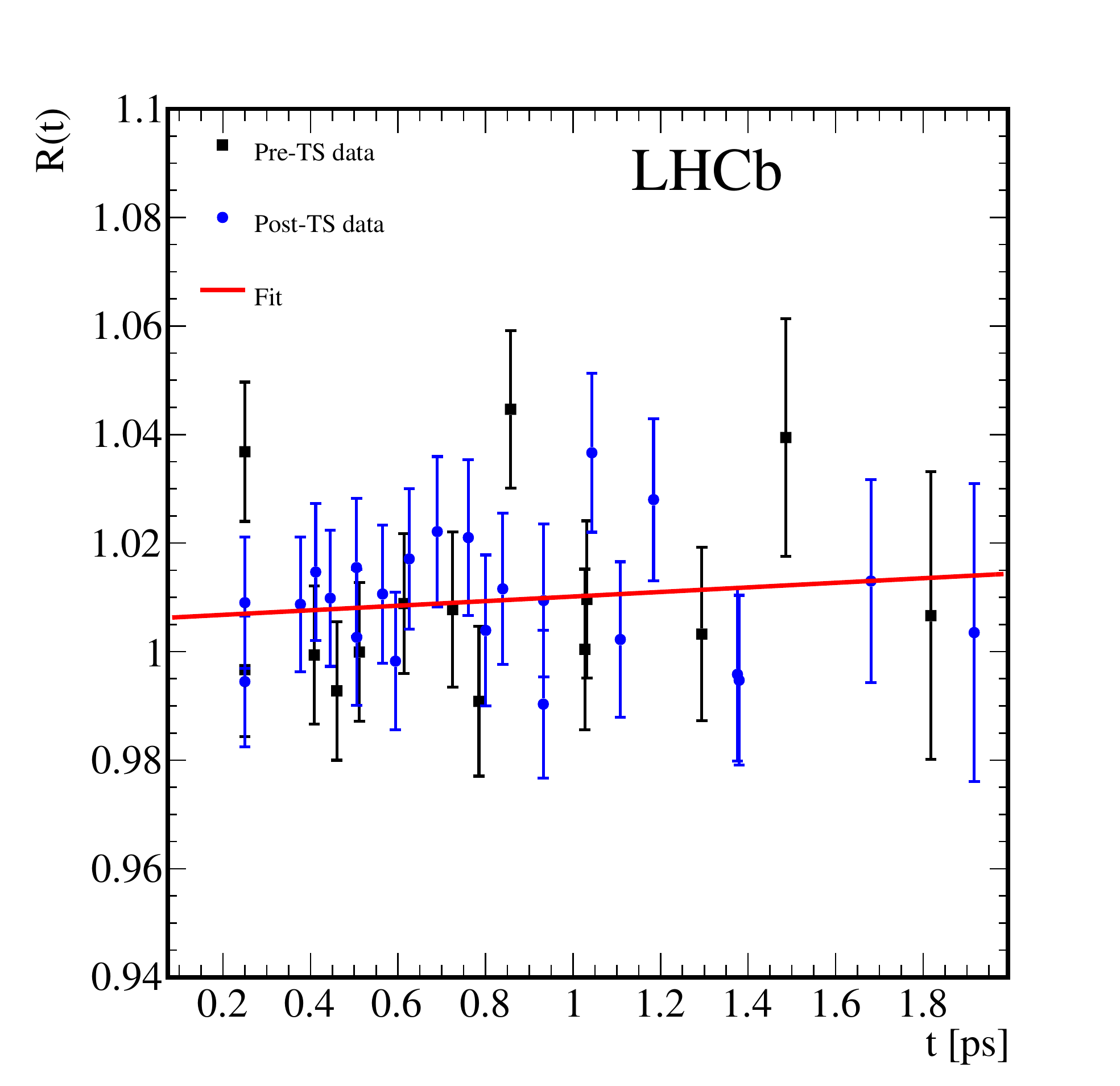}
\caption{Evolution of the of \Dz-\Dzb yield ratio with decay time for \dkkc  decays (on the left) and for \dpipic  decays (on the right) with the two data-taking periods shown with different symbols. The line shows the linear fit to extract \agamma~for the alternative method.}
\label{fig:ratio}
\end{figure}

An alternative method is used to perform \agamma~measurement on the
same data samples.  \agamma~is evaluated by the measurement of the ratio of \Dz and \Dzb yields
which can be written as
$$
R\left(t,t+\Delta t \right)\approx \frac{N_{\bar{ \Dznospace }}}{N_{ \Dznospace }} \left(1+\frac{2 \agamma}{\tau_{KK}} t \right) \frac{1-e^{\delta t/\tau_{\bar{ \Dznospace }}}}{1-e^{\delta t/\tau_{\Dznospace }}}
$$
where $\tau_{KK}=\tau_{K\pi}/(1+y_{CP})$ is taken from the world average value of this quantities.
By  simultaneous unbinned maximum likelihood fits to the \Dz mass, $\Delta m$
and  $\chi^2_{IP}$ one can extract the yields in each time bin.  
The evolution of the \Dz \Dzb yield ratio for the $K^-K^+$ and $\pi^-\pi^+$ samples are shown in Fig. \ref{fig:ratio}.

The two methods are tested in many simplified simulated experiments. 
From several cross checks no dependence or bias, e.g. on \Dz kinematics or multi-PV events,
are observed.

Several sources of systematics are considered, the main ones are due to the
decay-time acceptance correction and due to the background description. 
The first is assessed by testing the sensitivity 
to artificial biases applied to the per-event acceptance function.
The second is assessed through pseudo-experiments with varied background
levels and varied generated distributions while leaving the fit model
unchanged.
The systematics contributions are summarized in Table \ref{tab:sys_summary}.

The results of the four subsets are found to be in agreement with each other
for the nominal fit and the \agamma~measurements from the two methods yield 
consistent results.
The resulting value of \agamma~for the two final states evaluated by the main method is \cite{agamma}
\begin{eqnarray}
\agamma^{KK} & = & (-0.35\pm0.62_{stat}\pm0.12_{syst})\times10^{-3}\nonumber\\
\agamma^{\pi\pi} & = & (0.33\pm1.06_{stat}\pm0.14_{syst})\times10^{-3}.\nonumber\\
\end{eqnarray}

 The results for both final states are consistent with zero and hence no evidence of \CP violation is obtained.
They are also in agreement with the current world average.

\begin{table}[t]
\begin{center}
\begin{tabular}{|l|cc|}  
\hline
\hline
Source                               & $\agamma^{KK}$ & $\agamma^{\pi\pi}$\\
\hline
Part. rec. backgrounds                & $\pm0.02$ & $\pm0.00$ \\
Charm from B decays             & $\pm0.07$ & $\pm0.07$ \\
Other backgrounds                     & $\pm0.02$ & $\pm0.04$ \\
Acceptance function                   & $\pm0.09$ & $\pm0.11$ \\
\hline
Total systematic uncertainty          & $\pm0.12$ & $\pm0.14$ \\
\hline
Total statistical uncertainty         & $\pm0.62$ & $\pm1.06$ \\
\hline
\hline
\end{tabular}
\caption{Summary of systematic uncertainties, given as multiples of $10^{-3}$.}
\label{tab:sys_summary}
\end{center}
\end{table}

\section{The Heavy Flavour Averaging Group averages}
The Heavy Flavour Averaging Group (HFAG) \cite{hfag} provides averages for 
heavy flavour quantities. In particular, the  Charm Physics sub-group provides 
averages for the \Dz-\Dzb mixing and \CP violation parameters by combining measurements 
from different experiments. 

\begin{figure}[!tb]
\centering
\includegraphics[height=2.8in]{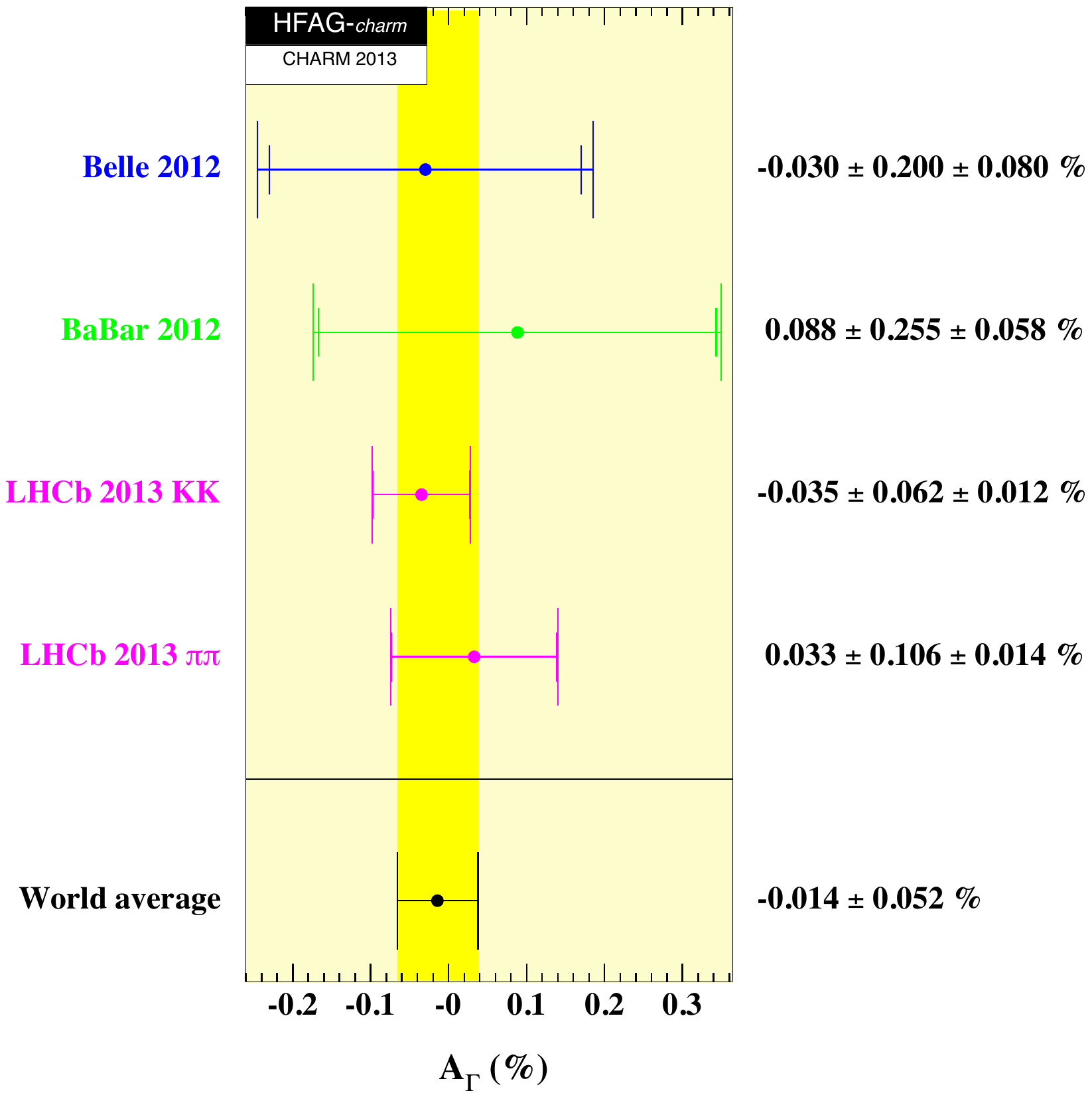}
\caption{World average value of \agamma~as calculated from \dkkc and \dpipic decays.}
\label{fig:agamma}
\end{figure}

The new world average of \agamma, shown in Fig.~\ref{fig:agamma}, is $-0.014\pm 0.052 \%$. Thanks to the contribution of 
the latest measurement  at per mille level by \lhcb, its accuracy has been increased by more than a factor 3.

\begin{figure}[!tb]
\centering
\includegraphics[height=3in]{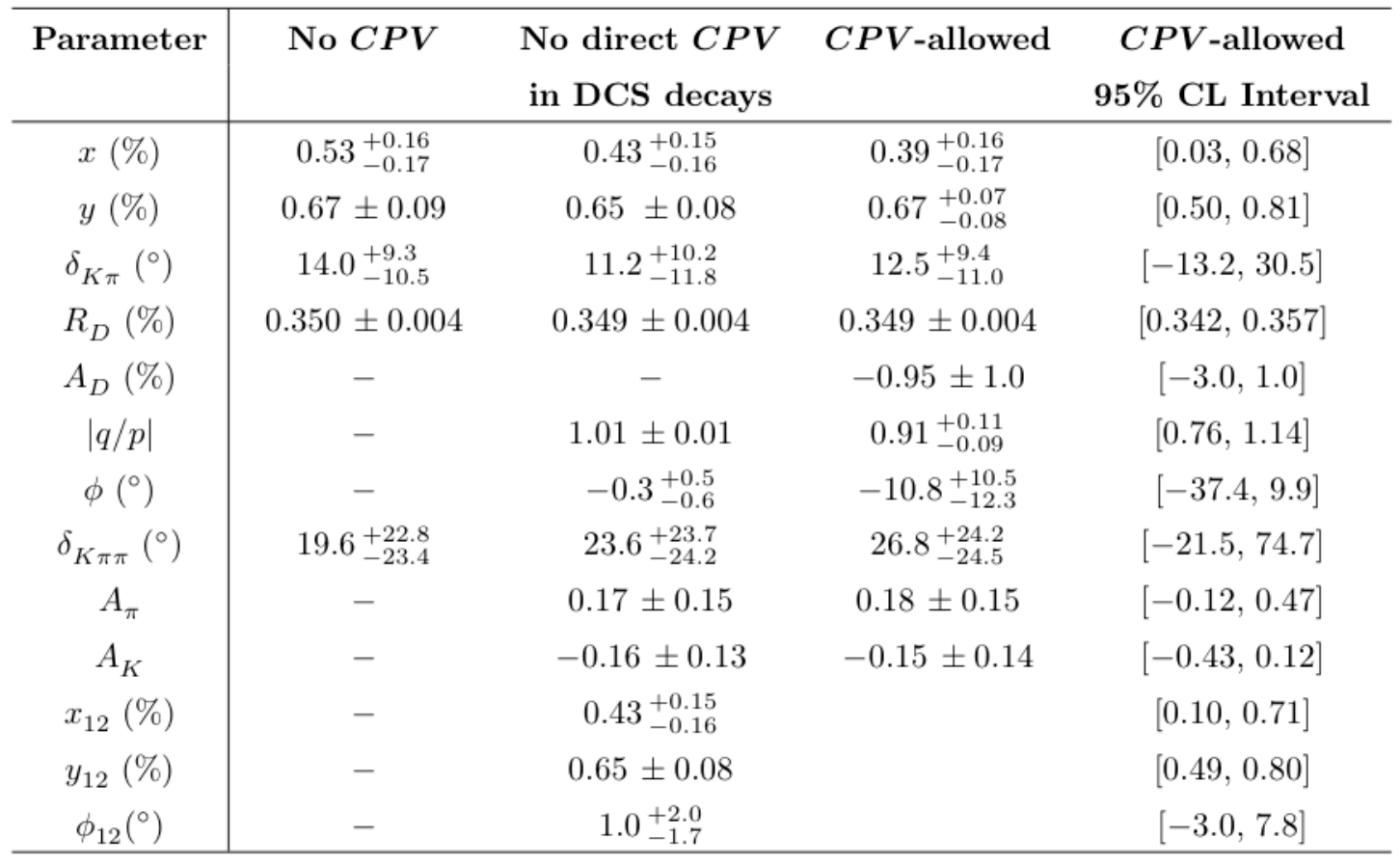}
\caption{Results of the mixing and \CP violating parameters by the global fit for different assumptions concerning \CP violation.}
\label{fig:table}
\end{figure}

In addition, HFAG uses a global $\chi^2$-based fit of all 41 experimental observables
from \belle, \babar, \cdf, \cleo and \lhcb experiments 
to extract the theoretical parameters describing the mixing and \CP violation in the charm sector. 
Correlations among observables are accounted for by using covariance matrices provided by the
experimental collaborations. Errors are assumed to be Gaussian, and systematic uncertainties
among different experiments are assumed uncorrelated unless specific correlations
have been identified.
The fit is performed considering three different \CP violation assumptions: no \CP violation in the charm sector, 
allowing indirect \CP asymmetry contribution but no direct \CP violation, and allowing both direct 
and indirect \CP violation. 
The results presented include the latest measurements of the mixing parameters and search of \CP violation in the double-Cabibbo suppressed 2-body decays at \lhcb \cite{agamma,lhcbmixing2}, and the latest mixing measurement at \cdf \cite{cdfmixing}.
All fit results are listed in Fig. \ref{fig:table}. 

\begin{figure}[!tb]
\centering
\includegraphics[height=2.4in]{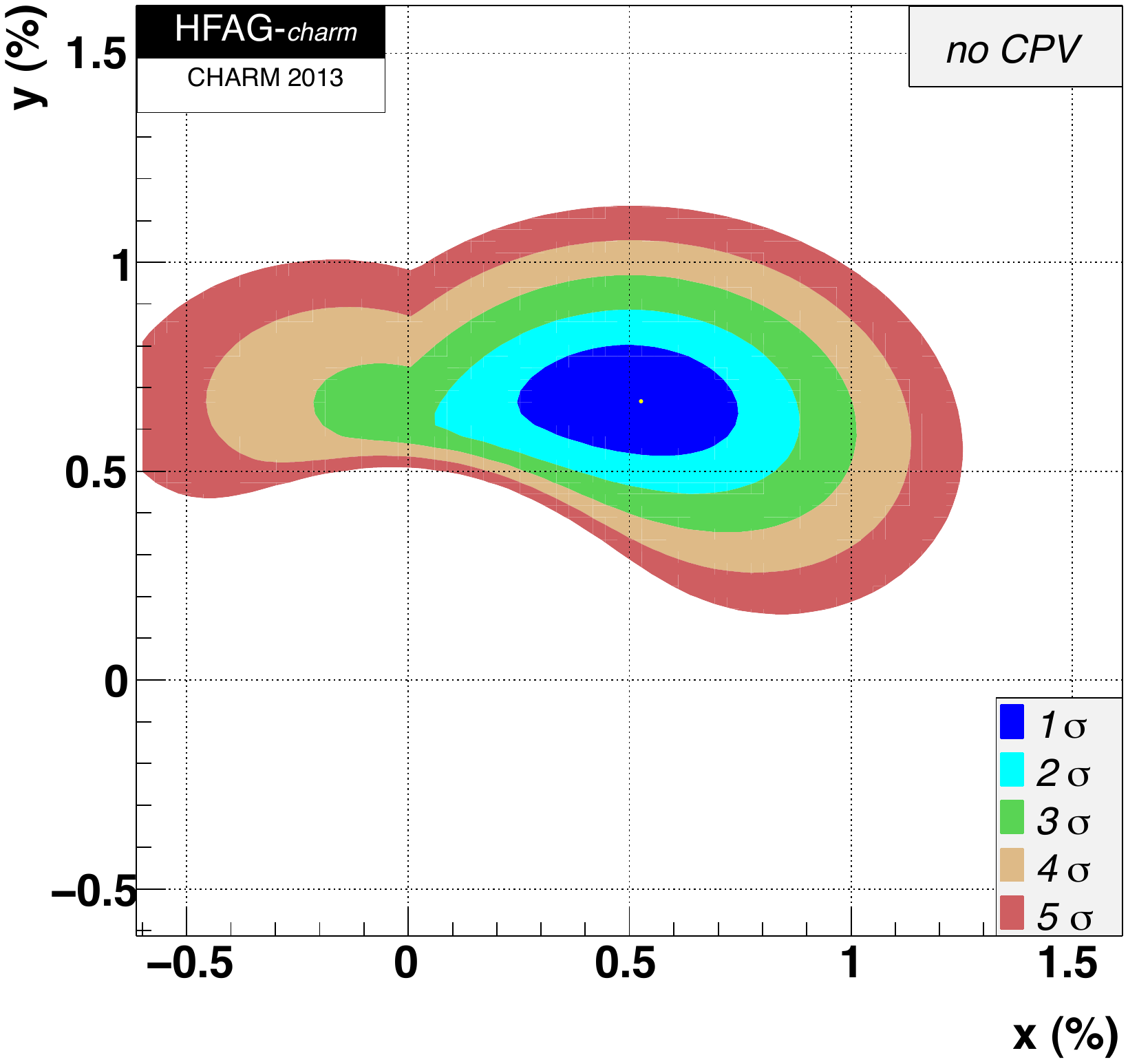}
\caption{Two-dimensional contours for the mixing parameters (x, y), for the no \CP violation assumption.}
\label{fig:xy}
\end{figure}

\begin{figure}[!tb]
\centering
\includegraphics[height=2.5in]{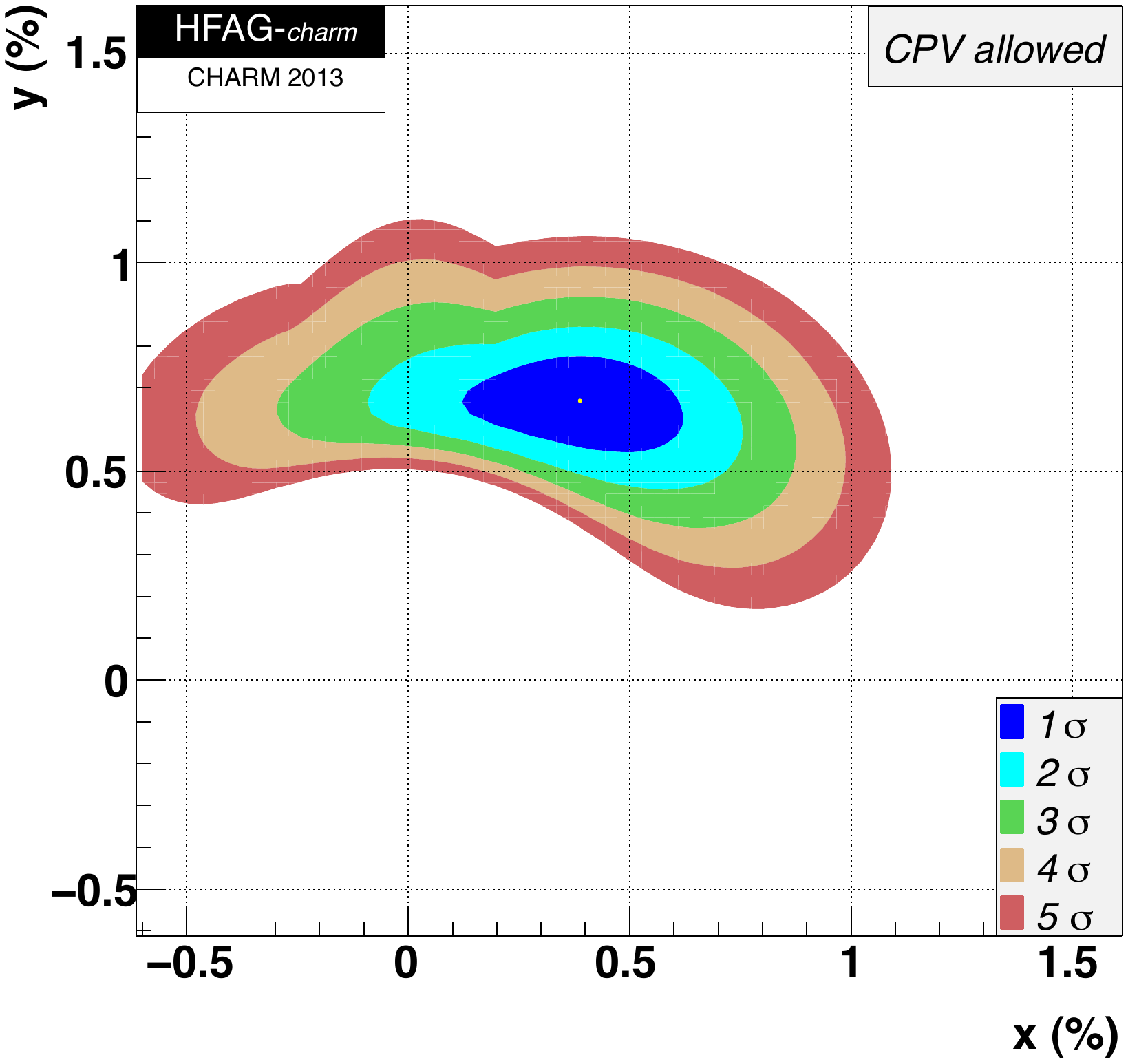}
\includegraphics[height=2.5in]{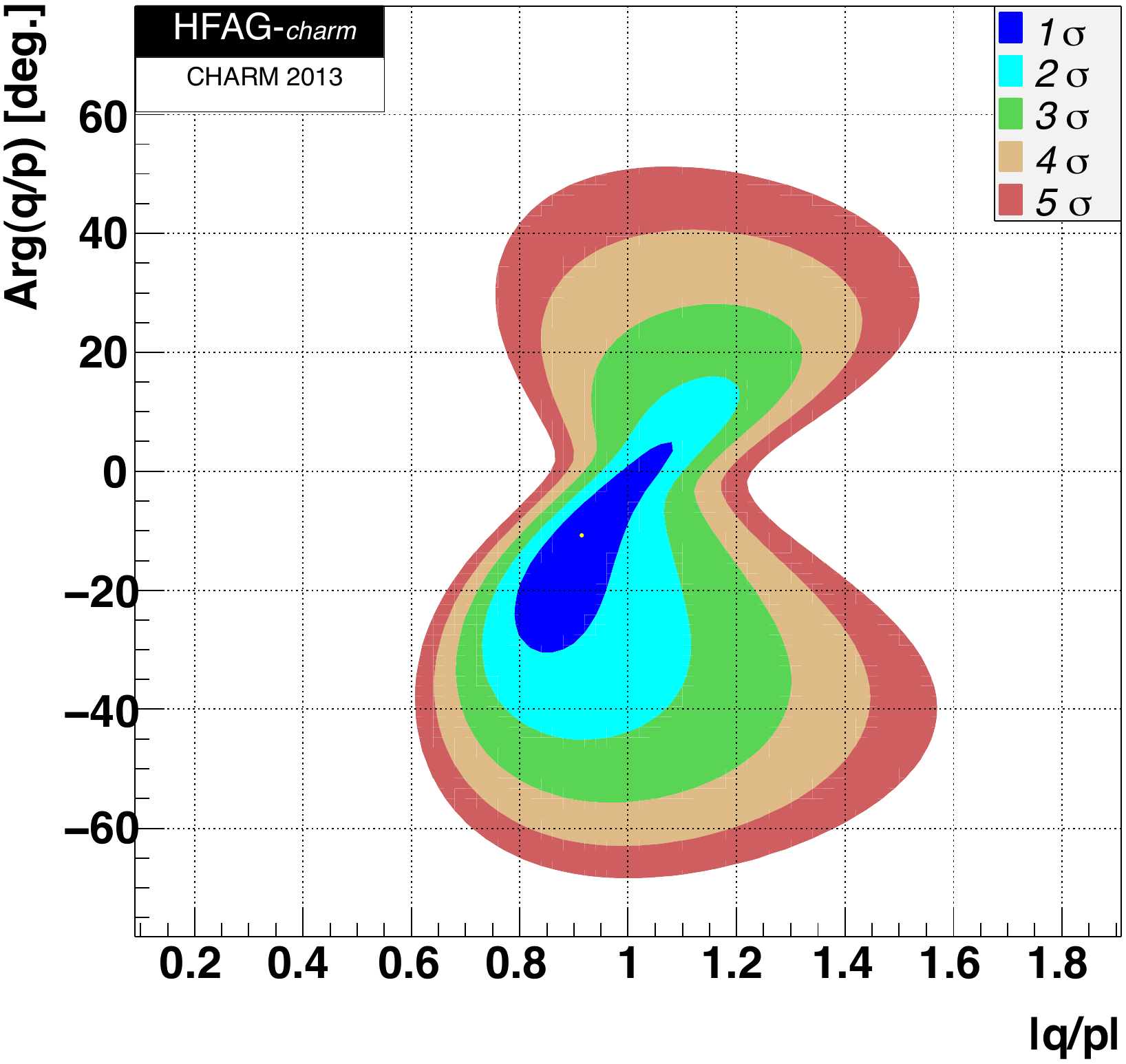}
\caption{Two-dimensional contours for the parameters (x, y) (left) and (|q/p|, $\phi$)
(right), allowing for \CP violation.}
\label{fig:qp}
\end{figure}

Confidence contour plots in ($x$, $y$) and ($|q/p|$, $\phi$ ) planes are obtained by letting, for any point in the two-dimensional plane, all
other fitted parameters take their preferred values. 
The resulting 1$\sigma$-5$\sigma$ contours are shown in Fig. \ref{fig:xy} for the \CP-conserving case and in Fig. \ref{fig:qp} for the \CP violation case.
In the latter fit, the no-mixing point $(x,y)=(0,0)$ is excluded at more than 12~$\sigma$. 
The no \CP violation point $(|q/p|,\phi)=(1,0)$  is within 1$\sigma$ with a confidence level (C.L.) at 48\%.
Thus the data are consistent with \CP conservation in mixing and in the interference between mixing and decay.

\begin{figure}[!tb]
\centering
\includegraphics[height=2.7in]{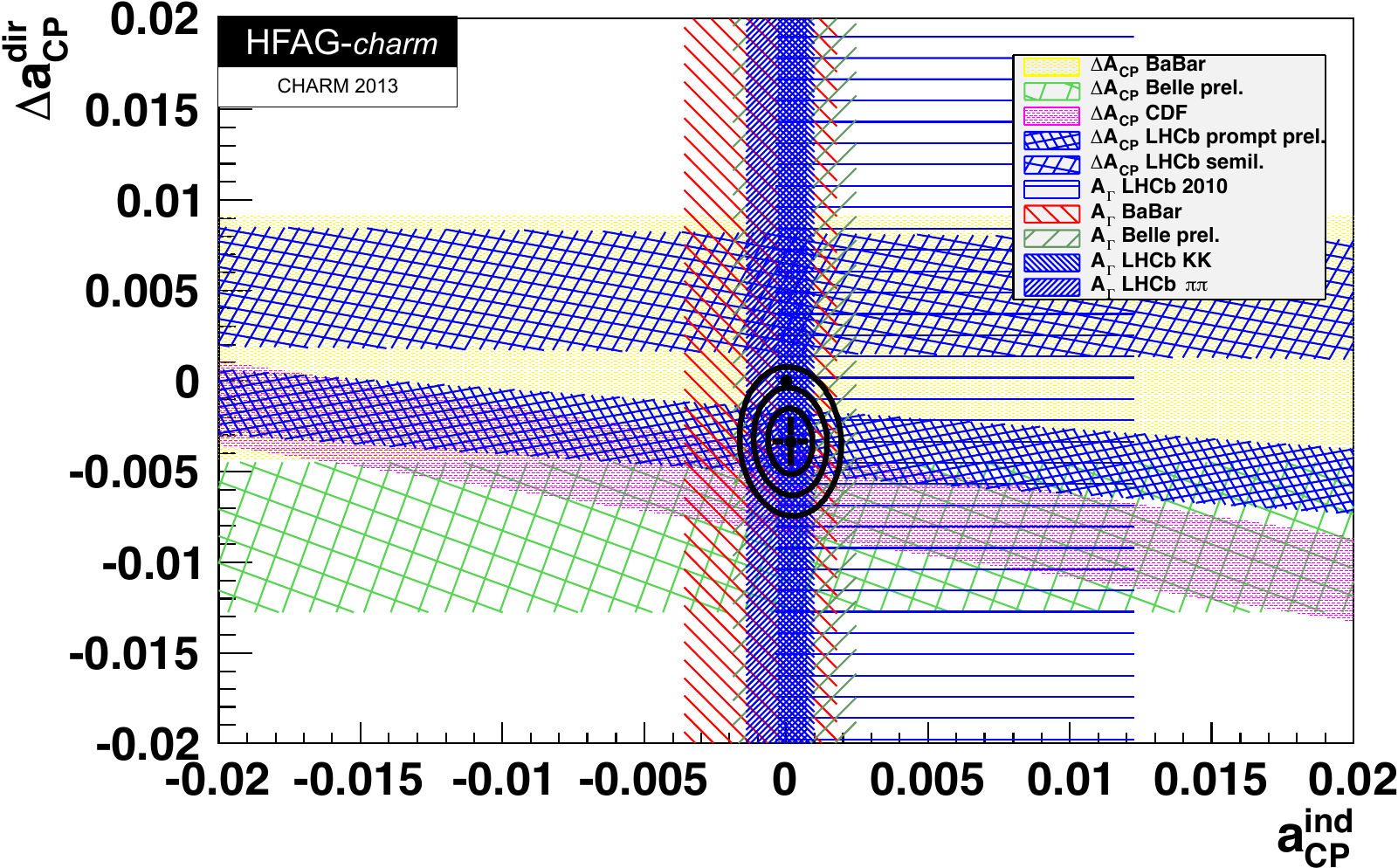}
\caption{The combination plot shows the indirect asymmetry versus the direct contribution. The bands represent $\pm 1 \sigma$ intervals.  The point of no CP violation (0,0) is shown as a filled circle, and two-dimensional 68\% C.L., 95\% C.L., and 99.7\% C.L. regions are plotted as ellipses with the best fit value as a cross indicating the one-dimensional uncertainties in their centre.}
\label{fig:directindirect}
\end{figure}

A nice way to visualize the status of search of \CP violation in the charm sector
is plotting the direct \CP asymmetry by Eq. \ref{eqdir} versus the indirect \CP asymmetry
defined by Eq. \ref{eqindir}. The direct contribution is mainly determined by the  $\Delta A_{CP}$
measurements \cite{acpfirst,acpcdf,acpsecond,acpsemilept}. 
The measurements of \agamma~determines the indirect contribution, the
contribution due to the direct \CP violation to \agamma~measurement is neglected.
The results are shown in Fig. \ref{fig:directindirect}.
The new world averages are  $a_{\CP}^{dir}=(0.015\pm 0.052) \%$ and $a_{\CP}^{ind}=(-0.033\pm 0.120)\%$.
These values are consistent with no \CP violation at 2.0\% confident level.

\section{Conclusion}
The search for indirect \CP violation in 2-body decays at the B-factories and at \lhcb are
consistent with \CP conservation.
The \agamma~measurements at \lhcb for the two \CP eigenstates are in agreement with each other and 
compatible with zero, as predicted by the SM at this level of precision.
HFAG combination excludes the no mixing hypothesis at more than 12~$\sigma$.
The  search of direct and indirect \CP violation in the  charm sector is consistent with
no \CP violation at 2.0\% confident level.
The next measurements at \lhcb, and later from Belle II and \lhcb upgrade will allow
to deploy deeply the \CP violation search in the charm sector.


\end{document}